\documentclass{egpubl}

\makeatletter
\def\JournalSubmission{%
\EGpagenumber%
\copyrightTextTitPag{\p@journalSubmissionText{(\number\month/\number\year)}}
\copyrightTextRunPag{\p@journalSubmissionText{(\number\month/\number\year)}}
\def\ps@titlepage{\let\@mkboth\@gobbletwo
 \def\@oddhead{\raisebox{\z@}[8pt][1pt]{\parbox{\textwidth}{\small
	~
   \hfill
 }}}%
 \def\@oddfoot{{\tiny\raisebox{\z@}[8pt][1pt]{\parbox[t]{18pc}{\sloppy
  \p@copyrightTextTitPag}}}\hfill}
 \let\@evenhead=\@oddhead
 \let\@evenfoot=\@oddfoot
 \let\sectionmark=\EmptySectionmark
 \let\subsectionmark=\EmptySubsectionmark
}}
\renewcommand\p@journalSubmissionText[1]{%
}%
\makeatother

 \JournalSubmission    % uncomment for submission to Computer Graphics Forum
\usepackage{graphicx}
\usepackage{microtype}
\DeclareGraphicsExtensions{.pdf,.jpg,.png}
\usepackage{t1enc,dfadobe}

\usepackage{cite}
\usepackage{amsmath}
\usepackage{mathptmx}
\usepackage{color}
\definecolor{darkred}{rgb}{0.5,0.0,0.0}
\definecolor{darkgreen}{rgb}{0.0,0.5,0.0}
\newcommand{\ti}[1]{\textcolor{darkred}{#1}}
\newcommand{\added}[1]{\textcolor{darkgreen}{#1}}
\renewcommand{\ti}[1]{#1}
\renewcommand{\added}[1]{#1}

\title{\mbox{Interactive Illustrative Line Styles and} \mbox{Line Style Transfer Functions for Flow Visualization}}

\newcommand{\myauthorshort}{M.H. Everts et al.}
\newcommand{\myauthorsimple}{Maarten H. Everts, Henk Bekker, Jos B.T.M. Roerdink, and Tobias Isenberg}
\newcommand{\myauthor}{\href{http://www.cs.rug.nl/svcg/People/MaartenEverts}{Maarten H. Everts},\textsuperscript{1} \href{http://www.cs.rug.nl/svcg/People/HenkBekker}{Henk Bekker},\textsuperscript{2} \href{http://www.cs.rug.nl/~roe/}{Jos B.T.M. Roerdink},\textsuperscript{2} and \href{http://www.cs.rug.nl/~isenberg/}{Tobias Isenberg}\textsuperscript{3}
\\
\textsuperscript{1}\,\href{http://www.tno.nl/}{TNO}, the Netherlands\hspace{7mm}\textsuperscript{2}\,\href{http://www.rug.nl/}{University of Groningen}, the Netherlands\hspace{7mm}\textsuperscript{3}\,\href{http://www.inria.fr/}{INRIA}, France}%

\newcommand{\mysubject}{Illustrative visualization, flow visualization, line visualization, line styles}
\author[\myauthorshort]
       {\myauthor
        \\
       }

\graphicspath{{figures_small/}{figures/}{./}}

\usepackage{hyperref}
\hypersetup{
  pdftitle={Interactive Illustrative Line Styles and Line Style Transfer Functions for Flow Visualization},
  pdfsubject={\mysubject},
  pdfauthor={\myauthorsimple},
  pdfcreator={LaTeX},
  plainpages=false,       % For problems with page referencing
  hypertexnames=false,    % For handling subfigures correctly
  bookmarksnumbered=false,% Include the section numbers in the list
  bookmarksopen=true,     % In the list, display highest level only
  bookmarksopenlevel=3,   % Display three levels of bookmarks
  pdfpagemode=UseNone,    % Show just the page
  pdfstartview=Fit,
  pdfborder={0 0 0},      % I don't want those silly boxes
  breaklinks=true,        % Allow breaking links
  colorlinks=false,       % Don't change coloring of links
  nesting=true}

\newcommand{\ie}{i.\,e.}

\newcommand{\eg}{e.\,g.}

\newcommand{\etAl}{et al.}

\renewcommand{\marginpar}[1]{}

\begin{document}

% \teaser{
%  \includegraphics[width=\linewidth]{eg_new}
%  \centering
%   \caption{New EG Logo}
% \label{fig:teaser}
% }

\maketitle

\begin{abstract}
  We present a flexible illustrative line style model for the visualization of streamline data. Our model partitions view-oriented line strips into parallel bands whose basic visual properties can be controlled independently. We thus extend previous line stylization techniques specifically for visualization purposes by allowing the parametrization of these bands based on the local line data attributes. \added{Moreover, our approach supports emphasis and abstraction by introducing line style transfer functions that map local line attribute values to complete line styles. With a flexible GPU implementation of this line style model we enable the interactive exploration of visual representations of streamlines.} We demonstrate the effectiveness of our model by applying it to 3D flow field datasets.  
\begin{classification} % according to http://www.acm.org/class/1998/
\CCScat{Computer Graphics}{I.3.3}{Picture/Image Generation}{Line and Curve Generation};
\CCScat{Computer Graphics}{I.3.m}{Miscellaneous}{Illustrative Visualization}
\end{classification}

\end{abstract}

%-------------------------------------------------------------------------
\section{Introduction}

% Intro flow
The flow of fluids and gases plays an important role in a wide variety of real-world phenomena. Examples include the aerodynamics of cars, the heat distribution in offices, and the airflow around falling ink droplets. Consequently, flow has been extensively studied, typically through three-dimensional simulations. These simulations yield large amounts of data containing information at multiple scales; for some applications the general structure of the flow is most relevant, for others the small local deviations are the subject of study. Visual representations of flow data help in understanding its behavior and over the years a large number of methods have been developed for this purpose. Initially, most flow visualization methods employed photorealistic rendering techniques, but later-on also methods that borrow principles from scientific illustration were developed.\vspace{-.05ex}

% WHAT, and WHY
Inspired by such illustrative visualization techniques \cite{Rautek:2008:IVN}, we present a flexible method for illustratively depicting streamlines generated from 3D vector fields. We achieve this flexibility by introducing a line style model whose parameters can be interactively manipulated, thus facilitating the interactive exploration of the parameter space of visual streamline representations. This allows the user to select and generate the representations that are most suitable for the data and communication goals at hand. In addition, we introduce \emph{linestyle transfer functions} that map scalar values derived from the 3D simulation to line styles. These line style transfer functions also allow us to actively control the introduced amount of abstraction, which is an important principle of illustrative visualization \cite{Rautek:2008:IVN}.

% HOW
In order to achieve flexible parametrization of line styles we generalize a previous illustrative approach for line visualization \cite{Everts:2009:DDH}, by subdividing the view-oriented line strips that represent the streamlines. These strips are split into bands that are arranged orthogonally to a line's direction, and whose shape, color, relative distance to the viewer, and width can be independently controlled. In addition, we allow these line parameters to individually depend on local data attributes such as temperature or velocity. The same local data attributes can also be used as input for the line transfer function, controlling which line style is being used for which part of a line. Together, the individual line parameter control and the line style transfer functions facilitate a flexible line-based illustrative visualization.

In summary, the contributions of this paper are a flexible line style model for use in scientific streamline visualizations (\autoref{sec:linestyle}), line style transfer functions to control the application of a specific style depending on an external parameter (\autoref{sec:transferfunction}), and a fast yet flexible implementation of this model on the GPU (\autoref{sec:implementation}). We demonstrate the power of our approach for a number of 3D flow datasets that exhibit complex flow patterns (\autoref{sec:results}) \ti{as well as through informal feedback from a domain expert (\autoref{sec:validation}). This technical report extends a previous short paper \cite{Everts:2011:ILF} and adds, in addition to a generally more detailed discussion, the use of directional color patterns (\autoref{sec:directional-color}), the concept of line style transfer functions (\autoref{sec:transferfunction}), as well as more results in \autoref{sec:results}.}%, as well as the evaluation of the resulting illustrative visualizations (\autoref{sec:validation}).}

\section{Related Work}
In this section we discuss related work in the fields of flow visualization and illustrative visualization.

\subsection{Flow Visualization}

Being one of the most fundamental subjects for visualization, a broad range of methods have been developed for the visualization of flow datasets. McLoughlin \etAl\ \cite{McLoughlin:2010:OTD} survey both flow visualization in general and integration-based, geometric flow visualization in particular. A key component of integration-based flow visualization methods is the use of geometric objects to depict the properties and structure of the flow. These objects are generated by integrating over the underlying velocity field---starting from a set of seed points.

Lines are the most widely used primitives for this purpose, and in the context of steady flow such trajectories are called streamlines, whereas for unsteady flow streak and pathlines are used. The challenge for the visualization of three-dimensional streamlines is overcoming perceptual problems. Simply rendering large numbers of lines can quickly lead to clutter and occlusion, not to mention the fact that the general thinness of a line makes it hard to convey depth and spatial relationships.

Solutions to deal with these perceptual challenges include careful placement of streamlines through seeding strategies (see \cite{McLoughlin:2010:OTD} for an overview), illuminated streamlines \cite{Zockler:1996:IV3}, and shaded tubes or ribbons \cite{Ueng:1996:ESS}. Particularly relevant for this paper are the approaches that employ textured view-oriented triangle strips to mimic shaded tubes \cite{Schussman:2002:SSO,Stoll:2005:VSL}. Such shaded primitives need to have a certain width for the depth perception to work and have only a limited number of visual variables to convey additional information: typically only width and color, although textures can also be used to convey information about the flow \cite{Stoll:2005:VSL}. In terms of flexibility in controlling the appearance of flow streamlines, the approach by Shen \etAl \cite{Shen:2004:IVT} that uses 3D flow textures is relevant.

\subsection{Illustrative Visualization and Line Stylization}

Illustrative visualization methods \ti{\cite{Rautek:2008:IVN,Brambilla:2012:IFV,Isenberg:2015:SIV}} use and apply the principles of (scientific) illustrators to achieve the clarity and effectiveness found in traditional illustrations. Naturally, many of these methods employ techniques from the field of non-photorealistic rendering (NPR). NPR methods particularly relevant to our work aim to replicate line drawings and, for this purpose, support different line styles.

Dooley and Cohen \cite{Dooley:1990:AIL}, for example, use dashing for illustrating geometric models, whereas the difference vectors of Schlechtweg \etAl\ \cite{Schlechtweg:1998:STL} permit a larger palette of styles. Other line style parametrization methods include stroke texturing \cite{Northrup:2000:ASH,Kalnins:2002:WND,Kalnins:2003:CSS}, multi-resolution curves \cite{Finkelstein:1994:MC}, skeletal strokes \cite{Hsu:1993:SS,Hsu:1994:DAU}, and programmable line styles \cite{Isenberg:2006:GCS,Grabli:2010:PRL}. These NPR styles are typically applied to contour and feature lines of 3D objects, aim to replicate marks made by traditional tools, and---if used in an illustration---may carry a meaning (such as parts being hidden). In our work, in contrast, we use line styles to specifically visualize data properties of streamlines in a flow. \ti{For a more comprehensive overview of illustrative techniques in flow visualization we refer the reader to Brambilla \etAl's survey \cite{Brambilla:2012:IFV}.}

Two concepts from the field of illustrative visualization are important for our work. The first is halos \cite{Appel:1979:HLE,Stoll:2005:VSL,Tarini:2006:AOE,Everts:2009:DDH} that make objects (including lines) easier to discern from the background, thus improving depth perception. \added{The second is the use of transfer functions, well known from volume rendering \cite{Kniss:2002:MTF}, to display multiple styles in one visualization, \eg, as used for volumetric style transfer functions \cite{Bruckner:2007:STF,Gerl:2012:SAI}.} In the context of flow visualization \ti{\cite{Brambilla:2012:IFV}}, other illustrative methods related to our work include stroke- and painting-inspired visualizations of 2D flow fields \cite{Kirby:1999:VMD,Li:2008:ISP}, illustrative 3D volume rendering \cite{Svakhine:2005:IPI}, stylized streamlines \ti{\cite{Mattausch:2003:SIE,Li:2007:IBS,Lawonn:2012:ASV}, textured streamlines \cite{Jianu:2007:VSR}}, animated, dashed streamlines \cite{Laramee:2005:GFV} and dashtubes \cite{Fuhrmann:1998:RTF}\ti{, as well as illustrative stream ribbons and surfaces \cite{Born:2010:ISS,Hummel:2010:IIR,Chen:2011:IVF}. We make use of these concepts in creating our own approach of illustrative line styles and line style transfer functions}.

\section{Line Styles for Visualization}
\label{sec:linestyle}

As we have seen, line-based flow visualizations are problematic due to their limited number of visual variables as well as the occlusion that is introduced if more than a few line primitives are being used. We address these two major issues by introducing an extended line style model for visualization purposes \added{(this section) and the use of line style transfer functions (\autoref{sec:transferfunction}) based on our extended line model}.

\subsection{Line Partitioning Into Line Bands}

\begin{figure}
    \centering
    \includegraphics[width=0.9\columnwidth]{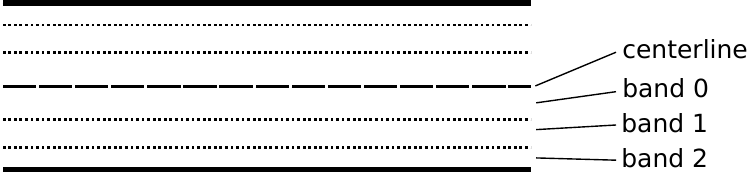}
    \caption{A view-oriented strip subdivided into a number of bands mirrored around the centerline.}
    \label{fig:mirroredbands}
\end{figure}

Such an extended illustrative line style model needs to increase the number of visual variables to allow the specification and parameterization of a variety of effects that can be flexibly used in visualization. For this purpose and inspired by halo-based line visualizations \cite{Everts:2009:DDH}, we partition the line strips that are used to render the data lines into several bands (\autoref{fig:mirroredbands}). By that separation we provide the granularity that is necessary to allow us to define visually distinguishable styles, each of the bands increasing the number of visual variables that can be controlled. These bands run parallel to the centerline and together define the visualization line style.

Specifically, we represent each line from the 3D dataset by a view-oriented line strip as done in many previous line-based rendering systems. This strip is subdivided into two mirrored sets of bands, one on each side of the line (\autoref{fig:mirroredbands}). The three visual properties that we control per band are color, width, and distance offset w.r.t.\ the viewer, each of which can be controlled independently. While the distance offset is not actually a \emph{visual} property, it has an effect when used as a depth-dependent halo \cite{Tarini:2006:AOE,Everts:2009:DDH}. In that case the halo line band is folded back, away from the viewer. The effect is that the \emph{visible} width of the halo depends on the difference in distance between two lines w.r.t.\ the viewer, improving the depth perception. Therefore, our extended linestyle model can be seen as a generalization of the depth-dependent line halo technique \cite{Everts:2009:DDH}.

\subsection{Local Attribute Mapping}

This basic line model allows us to specify a wide range of visual effects, notably by its capability to convey information about the data in the visualization by mapping data attributes such as temperature, pressure, etc. to a line's visual properties. Specifically, we control each line style band's color and width based on the value of local scalar line attributes by means of mapping functions.

For the color attribute, this mapping is encoded in conventional color maps that assign input values $\in [0,1]$ to RGB colors. We provide a selection of pre-defined color maps from which the user can choose a color map most suitable for that particular attribute type and the desired visual style. Similar to controlling the color of a band a user may adjust the width of a band to convey more information in the visualization. For example, mapping local velocity to band width yields wide bands where the velocity is high and thin bands where the velocity is low. To control this mapping, both a minimum and a maximum value can be set for the band width.

% Comments:
%\marginpar{Note: only linear right now, nonlinear can be encoded in color maps} 
%\marginpar{Note: this is only linear; non-linear possible with mapping functions below}

\subsection{Flexible Band Shapes}
\label{sec:bandshapes}

\begin{figure}
    \centering
    \includegraphics[width=0.8\columnwidth,height=0.45\columnwidth]{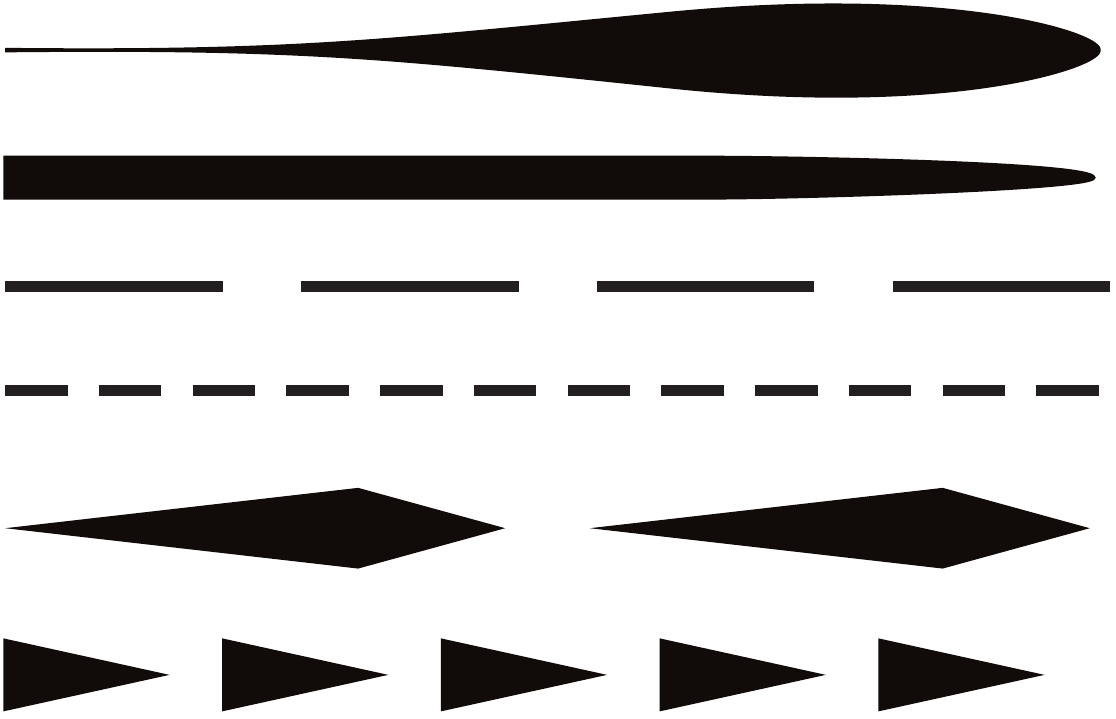}
    \caption{Example line shapes.}
    \label{fig:lineshapes}
\end{figure}

The control of the line width property of a band can also be used to create bands with repeating line shape patterns such as dashes, droplets, etc. (\eg, \autoref{fig:lineshapes}). Moreover, the local density (or frequency) of these repeating shape patterns can be used to provide additional means for conveying local flow properties; this is particularly useful for the velocity property. For this purpose we employ (1) a \emph{shape mapping function} and (2) a \emph{dedicated line shape attribute}. 
%\marginpar{Just notes: T: Just contemplating: What other properties could this be applied to? Only velocity? If so, Why? Maybe relation to speedlines in comics? Ref? M: well, theoretically one could calculate the cummulative sum of an attribute multiplied by the length of that line segment and use that: high values would yield elongated patterns, low values would yield short (high frequency) patterns. If the relation should be inversed, the cumsum of 1/attribute*length would probably give that effect. On problem is that such a attribute would be hard to describe and kind of unit-less.} 

A \emph{shape mapping function} maps a line data attribute ($\in [0,1]$) to the width ($\in [0,1]$) of the band at that point of a line. As such, it defines the shape of a band. We combine this mapping function with a \emph{dedicated line shape attribute}:\marginpar{M: replaced $s(s)$ with $s_x$, to avoid confusion with function} 
\begin{equation}
s_x = \frac{x}{l} \bmod 1,
\label{eq:shapeattribute}
\end{equation}
where $l$ defines the size of the shape pattern on the line and $x$ is a line attribute that is monotonically increasing along the line. The modulo operation ensures that the shape is repeated along the line. 

The choice of $x$ in \autoref{eq:shapeattribute} determines the local `density' of the patterns. For example, choosing the distance along the data line to the seed point results in constant size patterns.  However, choosing the integration time $t$ makes the frequency of the patterns depend on the local velocity of the field: high velocity will result in a lower frequency of patterns (\ie, elongated patterns), providing additional means for visualizing the local velocity.\marginpar{\small M: Removed (not yet existing) figure, sentence still makes sense?} 
%\autoref{fig:integrationtimepattern} illustrates\marginpar{M: maybe remove figure? Does it add much?} how this choice of the $x$ attribute gives additional means for visualizing the local velocity.    

%\begin{figure}
    %\centering
    %\includegraphics[width=0.9\columnwidth, height=0.3\columnwidth]{empty}
    %\caption{Illustration of how using integration time line attribute influences the size of repeating shape patterns.}
    %\label{fig:integrationtimepattern}
%\end{figure}

\begin{figure}
    \centering
    \includegraphics[width=0.9\columnwidth]{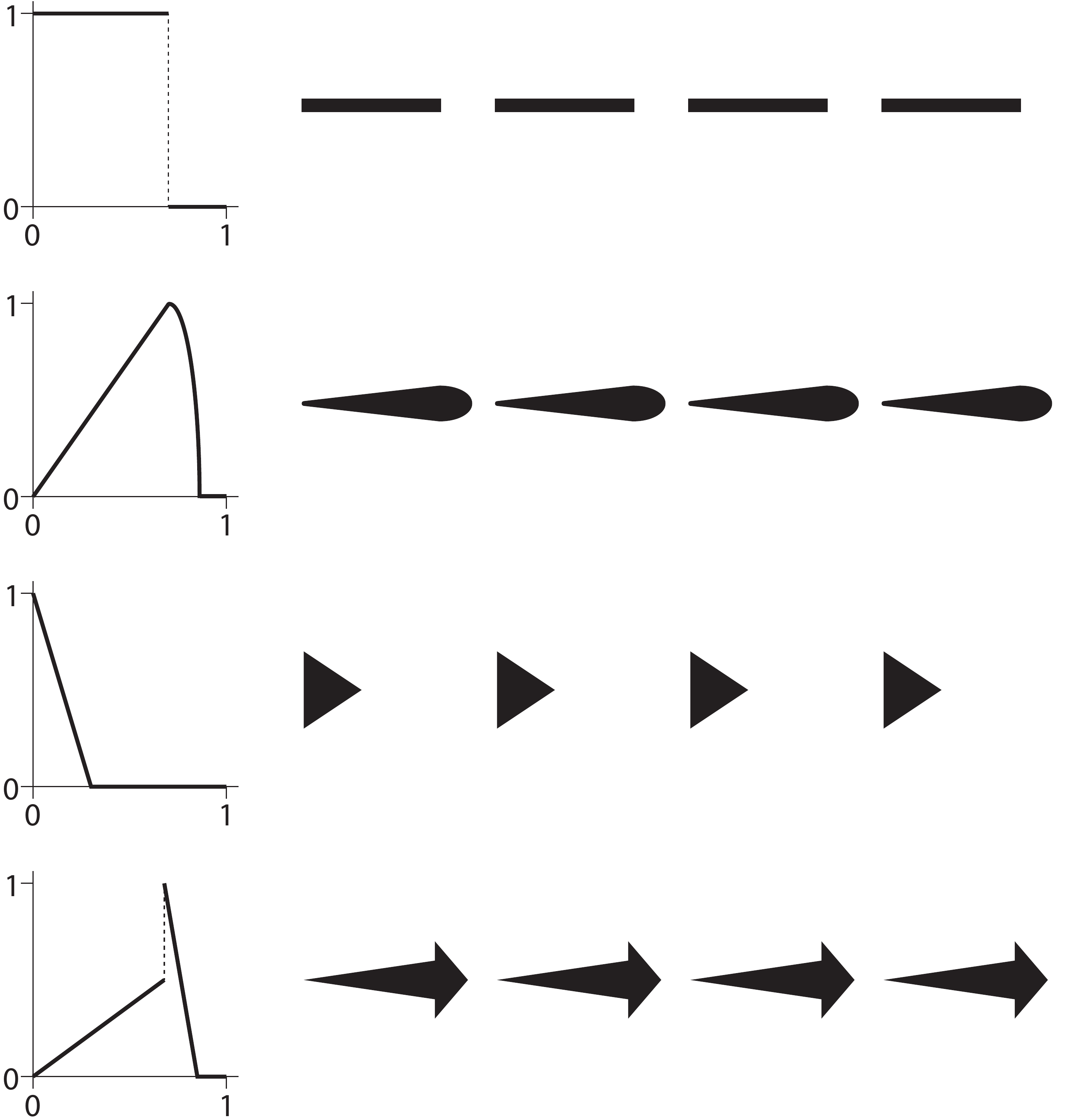}
    \caption{Shape mapping functions and corresponding line styles.}
    \label{fig:mappingfunctions}
\end{figure}

Together, the line shape attribute and the shape mapping function provide a flexible way to achieve a wide range of line shapes. \autoref{fig:mappingfunctions} shows a number of examples of mapping functions and illustrates how the mapping function influences the shape of a band and, thus, the line style.

\subsection{Directional Color Patterns}
\label{sec:directional-color}

\begin{figure}
    \centering
    \includegraphics[width=0.9\columnwidth,height=0.6\columnwidth]{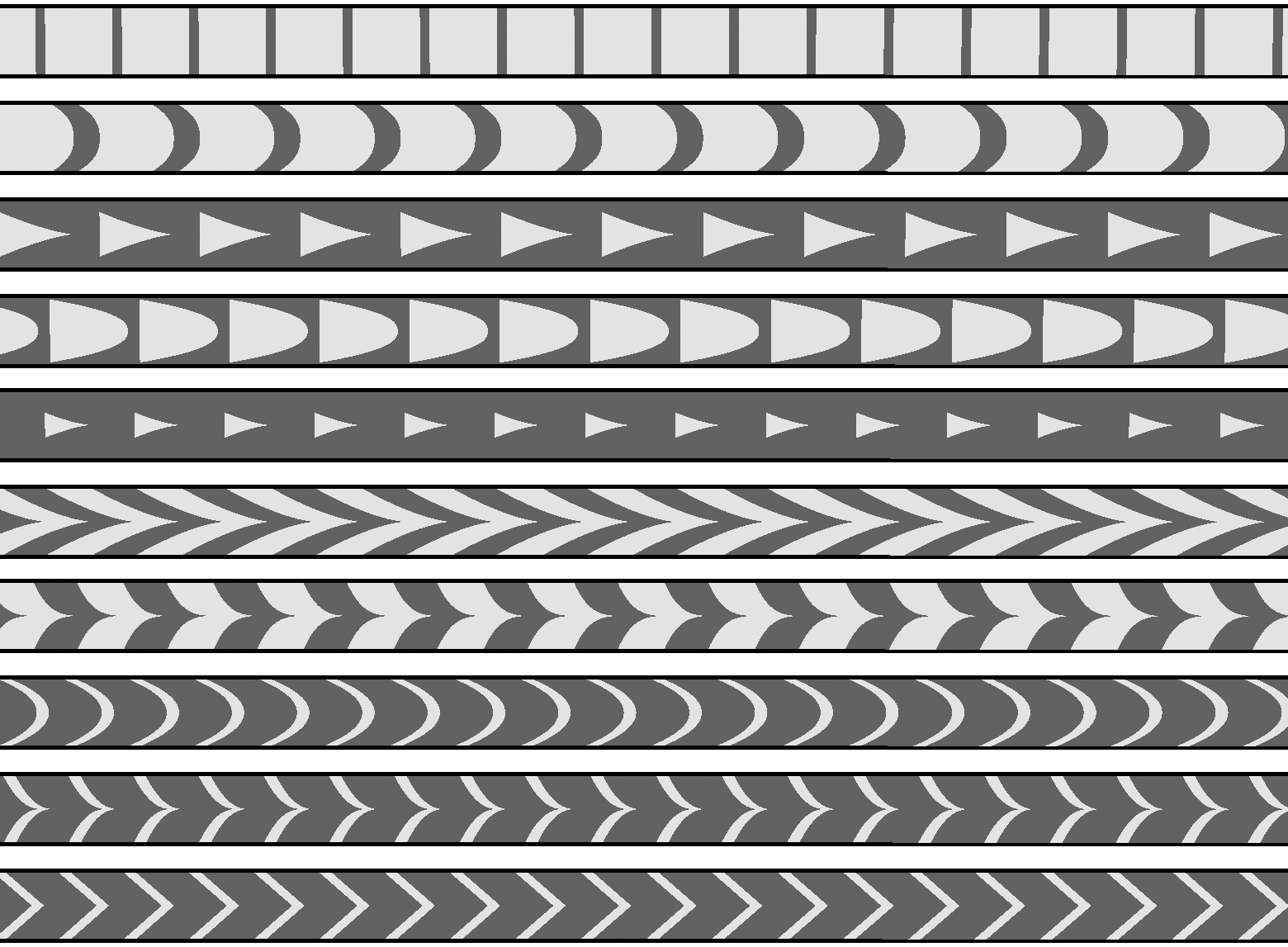}
    \caption{\protect\added{Examples of color patterns to apply to bands.}}
    \label{fig:colorpatterns}
\end{figure}

\added{The flexible band shapes (line band width mapping) affect the overall shape of the line as a whole.\marginpar{M: removed reference to centerline offset, might need rewording.} For example, an adjustment of the mapping of an inner band will always have an effect on the outer band layers. To provide a related visual effect without the influence of the overall line shape we provide---as the final aspect to our visualization line style model---\emph{directional color patterns} realized as procedural textures that control the line color of the bands. These textures can be employed to produce similar shapes and patterns (\eg, \autoref{fig:colorpatterns}) as the flexible band shapes but without affecting the line width. Similar to the band shapes, the directional color patterns can also convey additional information about the streamlines such as direction, velocity, and other parameters.}

\added{The approach we use to achieve this type of patterns is to apply a function that, based on the position on the band, determines which band color to use. For example, a decision function that captures the different aspects of the patterns illustrated in \autoref{fig:colorpatterns} is:
\begin{equation}
f_d(x,b) = \left(\left( \frac{x}{l} + a\,b^c \right) \bmod 1\right) - w,
\label{eq:decisionfunction}
\end{equation}
where $x$ is again a line attribute that is monotonically increasing along the line, defining the longitudinal position on the strip, $b \in [0,1]$ is the lateral position on the band, $l$ is the length of the repeating pattern, $a$ is the steepness of the slope, $c > 0$ defines the shape of the slope, and $w \in [0,1]$ determines the relative width of the colors. When $f_d$ is smaller than zero, one color is used, otherwise a second color. Similar to the band shape and centerline patterns, choosing $x$ to be the integration time in \autoref{eq:decisionfunction} allows the color pattern to convey velocity through the local frequency of the pattern.}

\subsection{The Extended Line Style Model for Visualization}

Together, the line bands with their color and width control, the means to parametrize the band width with mapping functions, and the directional color patterns extend the number of visual variables available for visualization of line data.  Most of these visual variables can be combined in one visualization and convey multiple aspects of the streamline data in one image.

\added{For example, consider a line style with the following settings for its two bands. On the center-most bands, the temperature is mapped to color via a color map. The resulting color is used as one of the colors in a directional color pattern to convey both velocity and directionality using another constant color. The second band (on both sides) is used as a contour line and is simply black. The width of both bands depends on the pressure. In such a visualization, the viewer can discern four different aspects of the flow field (besides the path of the streamlines): temperature, pressure, velocity, and direction---all aspects discernible in the mapping.}

\added{This combination of information comes at a price: the resulting image may be overwhelming and cluttered. However, we do not necessarily need to use all visual variables for all streamlines. One could imagine that, depending on the illustrative visualization goal, we would like to see extra information about the flow only for streamlines that have certain properties (\eg, local attribute values). To cater to this need, we introduce \emph{line style transfer functions} as described next.}

\section{Line Style Transfer Functions}
\label{sec:transferfunction}

\added{Similar to regular transfer functions in volume rendering \cite{Kniss:2002:MTF}, \emph{line style transfer functions} assign pre-defined line styles to line regions with certain local line attributes. Because we use our extended model for these line styles, the styles assigned to a attribute range are typically not static such as traditional (NPR) line styles but themselves show changes in the attribute values. Therefore, line style transfer functions allow us to visualize attribute values on two levels: attribute level ranges and changes within these ranges. Moreover, this line style transfer functions also facilitate the explicit control of abstraction in streamline visualization: by choosing a line style that maps fewer attributes using a less complex line style (which can be as simple as a single thin line, maybe with a dashing pattern) for regions of the dataset that are less important (as indicated by a given attribute), we can highlight important regions with a more elaborate style that encodes more attributes for a detailed visualization.}

\added{As an example of explicitly applying abstraction, let us consider a visualization user who is interested in the behavior of streamlines in high vorticity areas, but still would like to see the streamlines in the remainder as context. So he or she creates two line styles: one style with thick colormapped lines (to show temperature) combined with halos\marginpar{M: changed, ok?} and one with thin, dashed, light gray lines. The line style transfer function is then configured such that high vorticity values use the first line style and low vorticity values use the latter. Thus, low vorticity lines are thin and dashed so one can see through them, yet they provide context for the more pronounced high vorticity regions of the dataset.}

\added{Conceptually, a line style transfer function is a simple mapping from ranges of line attributes to line styles. As such, because the transfer function depends on local line attributes, one line in the visualization can have multiple line styles.\marginpar{M: removed transition stuff, there is no transition in the line style transfer function atm}}

%This means that we need to transition from one line style to the next for parts of a line that connect two attribute ranges with different line styles mapped to it. This transition\marginpar{M: there is no such transition, it's abrupt, from one to the other! To fix!} is facilitated through our banded line style model, so we can make the transition between associated bands (using empty bands if one style has less bands than the adjacent one).\marginpar{M: Need to think about this.} While some properties can be smoothly transitioned between \dots\marginpar{T: this transition needs to be described in more detail, say that some are smooth and others are not. how does this happen anyway? linearly between two adjacent vertices each with a attribute value belonging to a different line style? or suddenly in-between such two vertices? anything else about the transition?} Note though that the transition from one line style to the other is not necessarily smooth.

\section{Implementation}
\label{sec:implementation}

Several design decisions of the conceptual line model and transfer functions were driven by implementation considerations. More specifically, because we aim for the interactive exploration of line styles even when applied to large datasets, the line style model should be suitable for implementation in shaders on modern GPUs. Our implementation consists of two parts, each implemented in a different type of shader. The first is the generation of view-oriented triangle strips (geometry shader) and the second is the application of the line style to the strip (fragment shader).

% Generation of view-oriented triangle strips
The transformation of the input lines (stored in GPU memory) into view-oriented triangle strips is done each rendering pass in a geometry shader.  The width of these triangle strips is (pre-)calculated by multiplying a global scaling factor with the maximum of the line style widths. The width of a line style is calculated through a summation of the maximum widths of its bands. With the line strips in place as the `canvas' for the line style, the next step is to apply the style model. 

% Application of the line style
The actual application of the line style is done in a fragment shader. The main goal of this fragment shader is to decide which band of which line style should be applied to the fragment. To determine this it uses the position on the strip, the shape mapping functions, and the values of the relevant line attributes.  Then, based on the settings for that band, the color (either from a color map or from a color pattern) and the depth offset of the fragment can be determined. \marginpar{M: reworded, still ok?}

%\marginpar{T: be more precise here: which mapping functions, there are quite a few; maybe name them mathematically earlier?},
One additional aspect of our implementation is the use of templated shaders. The main reason for this is that the flexibility of our extended line style model yields a large number of options, which without templated shaders would result in a large number of expensive conditional statements in the shader. The templated shaders (implemented using the existing templating library Jinja2, see \href{http://jinja.pocoo.org/docs/}{http://jinja.pocoo.org/docs/}) allow us to flexibly include only the necessary shader code, based on the chosen style configurations. This approach has the additional benefit of making the shader used for rendering as small as possible. 
%\marginpar{T: ref? M: Can't find ref, but it's not ground-breaking, too simple to ref?} 

\section{Results}
\label{sec:results}

To illustrate the broad range of possible visual representations of lines that can be achieved with our line style model, we apply a number of different line styles to two sets of streamlines. The first set is generated from a snapshot of a numerical simulation of a heat driven cavity (Dataset 1), the other set is generated from a snapshot of a simulation of turbulent flow around a cube (Dataset 2). It is important to note that although the streamlines that we visualize here may give the impression of a steady flow, they are merely a visualization of the flow field at one particular time step.

\begin{figure}[tcb]
  \centering
 \includegraphics[width=\linewidth]{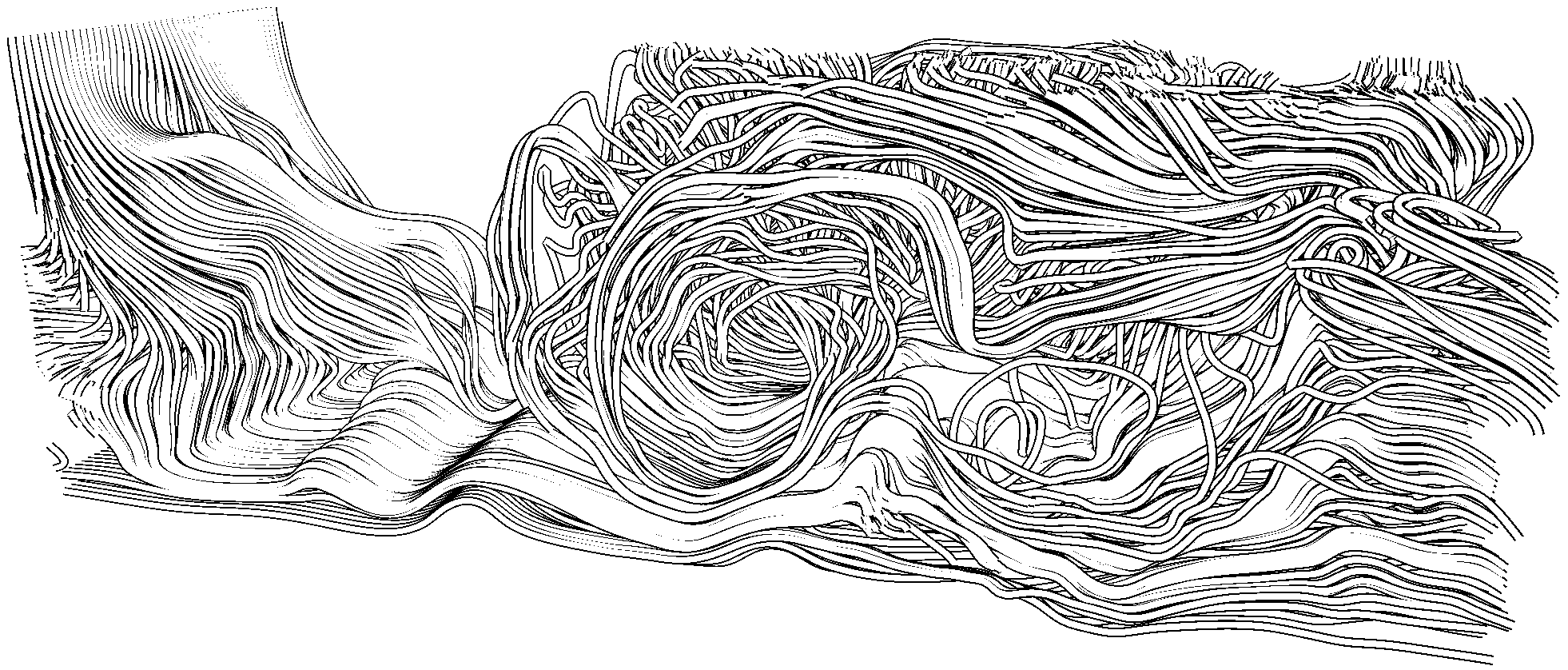}
  \caption{\label{fig:simplebw} A simple black-and-white line style applied to streamlines from dataset 1.}
\end{figure}

\begin{figure}
    \centering
    \includegraphics[width=\columnwidth]{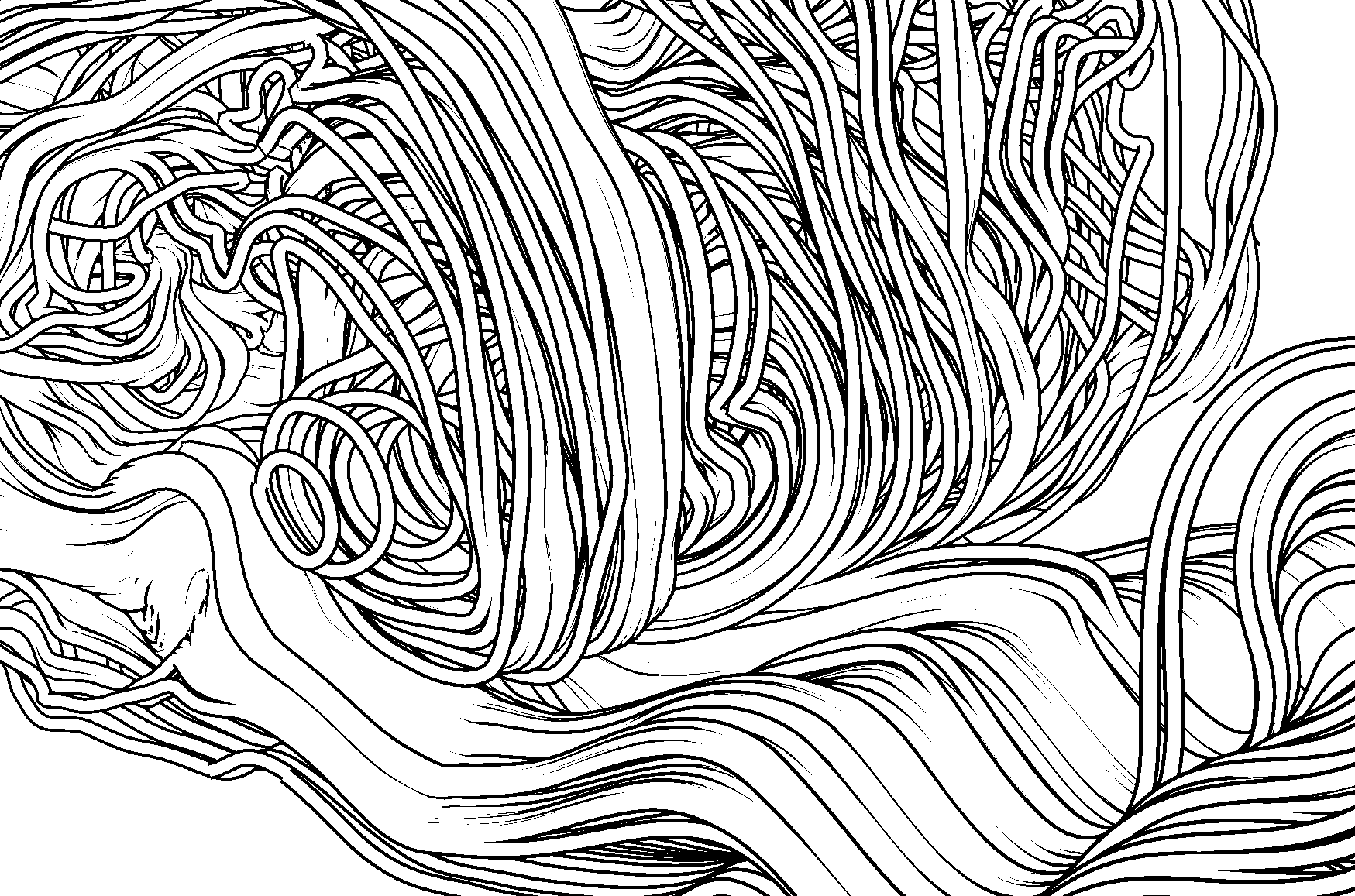}
    \caption{Close-up of streamlines with a simple black-and-white line style applied to them. Notice how the depth-dependent halo (contour) emphasizes collinear streamlines.}
    \label{fig:bw_closeup}
\end{figure}

We start with the application of a single simple black-and-white line style to Dataset 1 (See \autoref{fig:simplebw}). This line style has two bands. The center-most band is white and fairly wide, whereas the outer band is thin and black. In addition, this outer band acts as a depth-dependent halo, although in this case it can also be considered a depth-dependent contour. The first thing to notice in \autoref{fig:simplebw} is how, despite the fact that no color has been used, the spatial relationships of the lines are still clear. Also, the depth manipulation ensures that collinear streamlines (\eg, the laminar flow at the bottom) blend together, emphasizing such collinear structures and yielding a crisper visualization. The close-up of the same dataset in \autoref{fig:bw_closeup} illustrates this aspect further.

\begin{figure}
    \centering
    \includegraphics[width=\columnwidth]{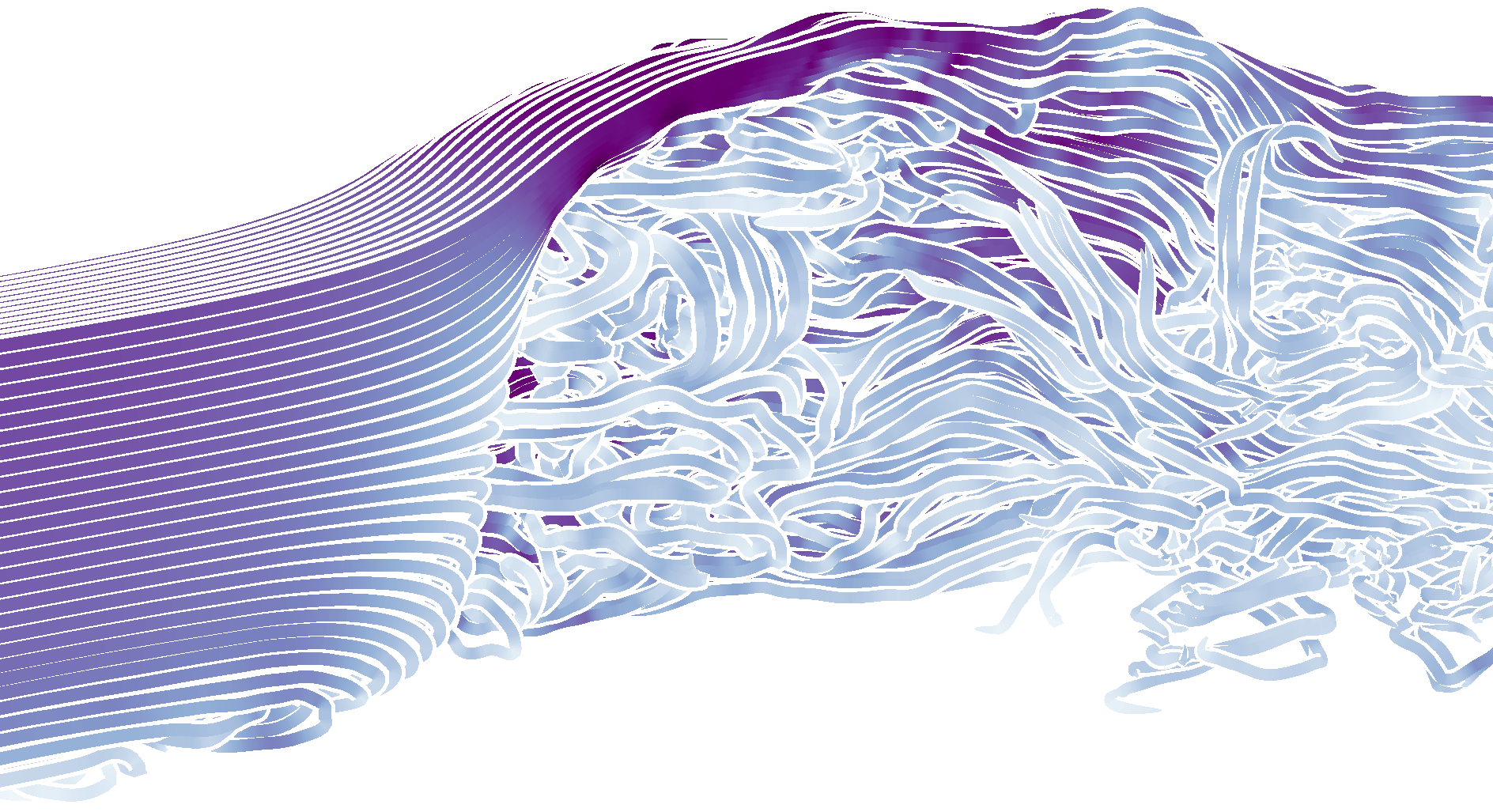}
    \caption{Streamlines depicting flow around a cube, colored with a blue-purple color map to show velocity, combined with white halos for depth perception.}
    \label{fig:onlycolormap}
\end{figure}

\begin{figure}
    \centering
    \includegraphics[width=\columnwidth]{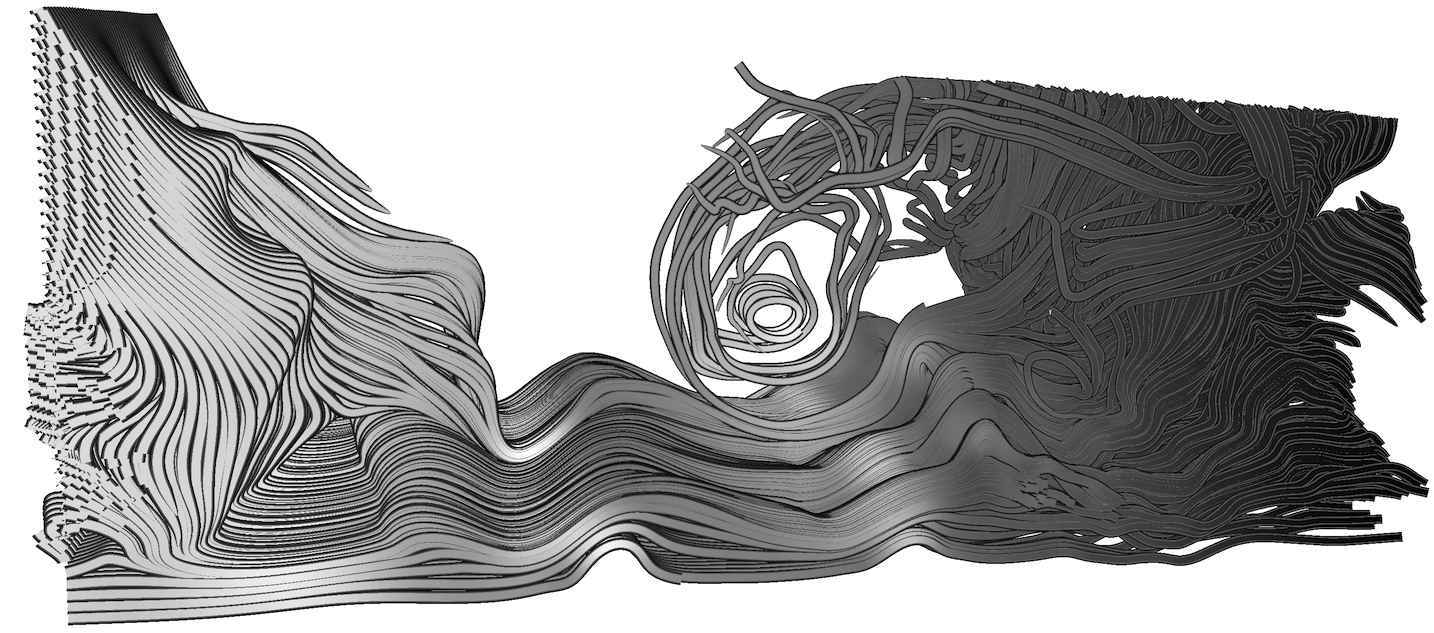}
    \caption{\protect\added{Gray scale color map applied to streamlines depicting their local temperature.}}
    \label{fig:grayscalecolormap}
\end{figure}

The next step is to employ the visual parameters that our line style model introduces to convey additional information about the flow. Figures \ref{fig:onlycolormap} \added{and \ref{fig:grayscalecolormap}} show the application of a color map to streamlines. In \autoref{fig:onlycolormap} a blue-purple color map is used to display velocity in Dataset 2; \added{and in \autoref{fig:grayscalecolormap} a grayscale color map is used to convey local temperature in Dataset 1}. Again, the halo allows us to omit shading and still have good depth perception, making direct application of color maps possible without a potential shading that affects the perception of the colors.

\begin{figure}
    \centering
    \includegraphics[width=\columnwidth]{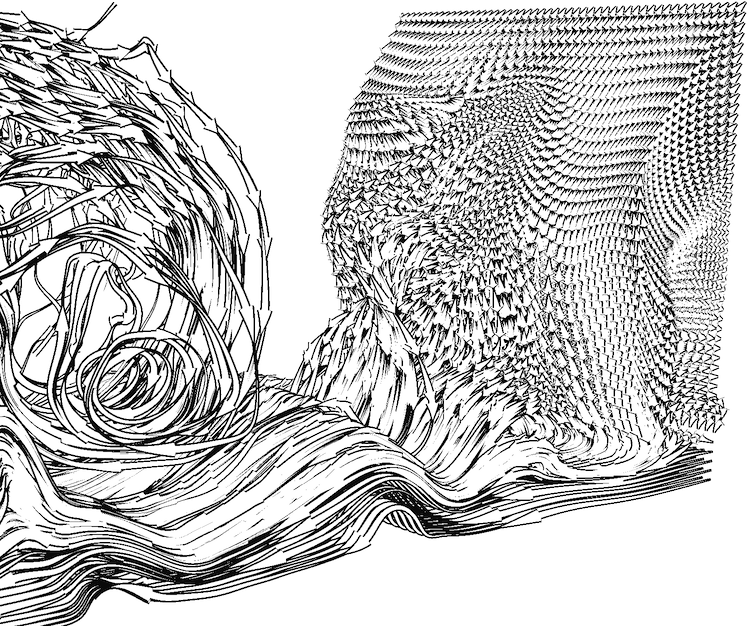}
    \caption{An arrow shape mapping function applied with a simple black-and-white style. The size of the arrow indicates velocity.}
    \label{fig:bwshape}
\end{figure}

\begin{figure}[tcb]
  \centering
 \includegraphics[width=\linewidth]{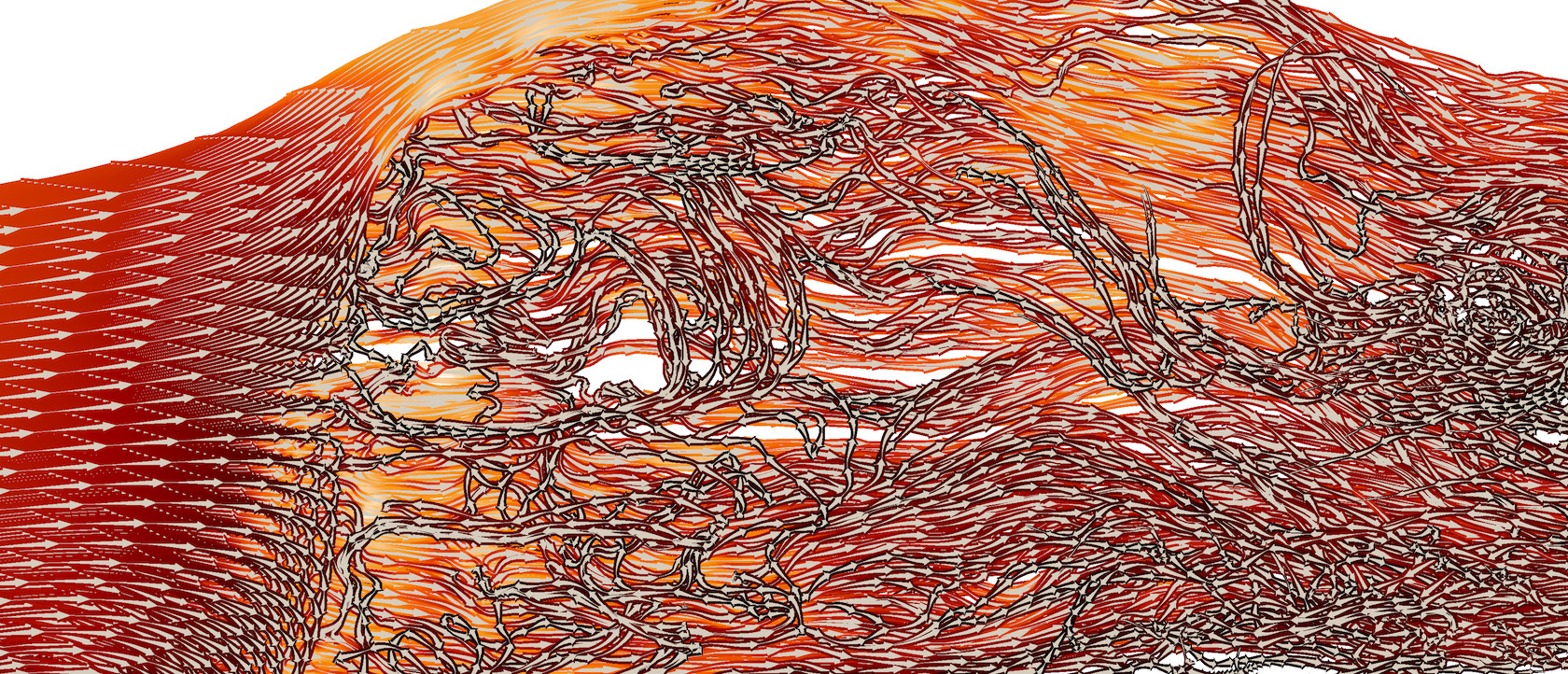}
  \caption{\label{fig:whitearrowcolormap} Streamlines depicted through light gray arrow shapes combined with a halos colored with a color map to depict velocity.}
\end{figure}

\begin{figure}
    \centering
    \includegraphics[width=\columnwidth]{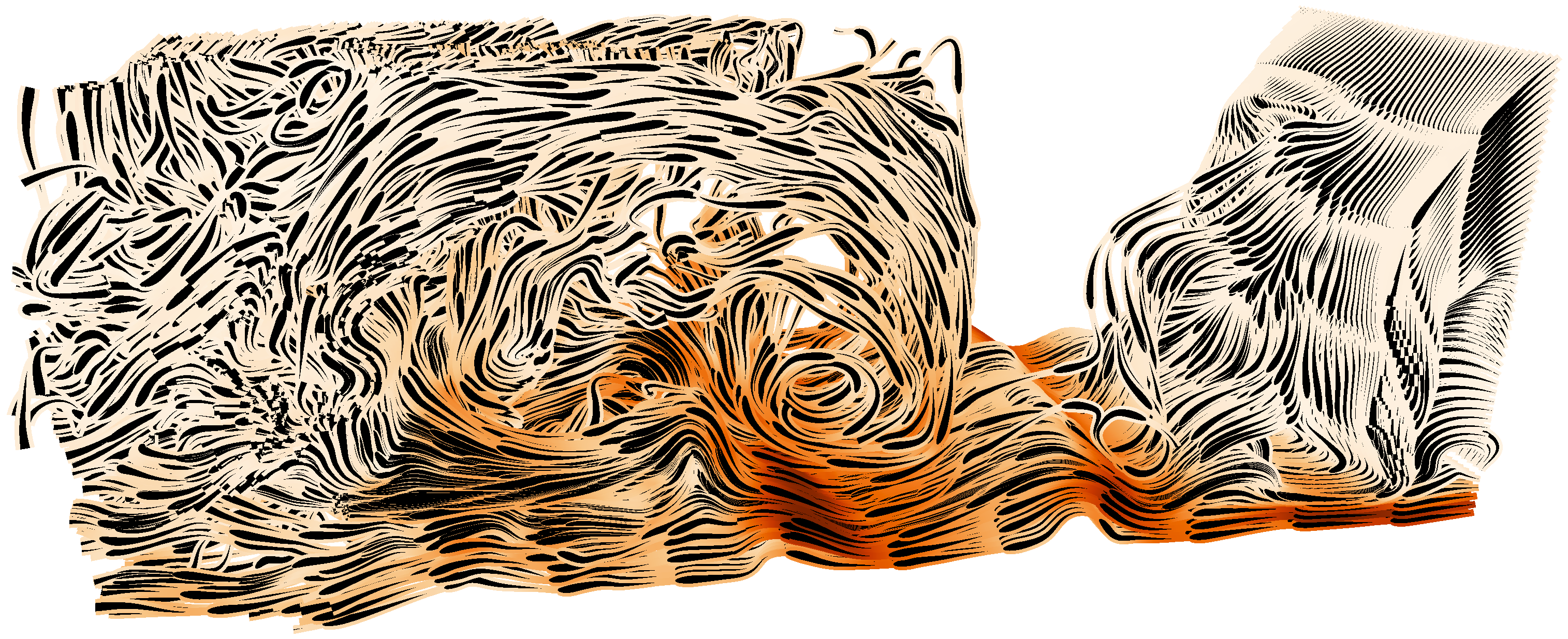}
    \caption{Streamlines depicted with tadpole-shaped, fixed size patterns, combined with a `halo' colored using a colormap (velocity).}
    \label{fig:blobbyshape01}
\end{figure}

Besides color maps, the other way our line style model can convey additional information is through the size and frequency of shape patterns. \autoref{fig:bwshape} illustrates how an arrow shape can be used to convey both direction and velocity in a black-and-white visualization. Shape and color maps can also be combined, as illustrated in \autoref{fig:whitearrowcolormap}, where light gray arrows are combined with a fairly wide halo to which a color map is applied. Again the size (length) of a pattern indicates the local velocity of the flow. Besides an indication of direction, the arrow shape also gives the visualization a certain feel of motion. A similar effect is achieved with the tadpole shape shown in \autoref{fig:blobbyshape01} where also a color-mapped halo is used, but with a constant shape length.

\begin{figure}
    \centering
    \includegraphics[width=\columnwidth]{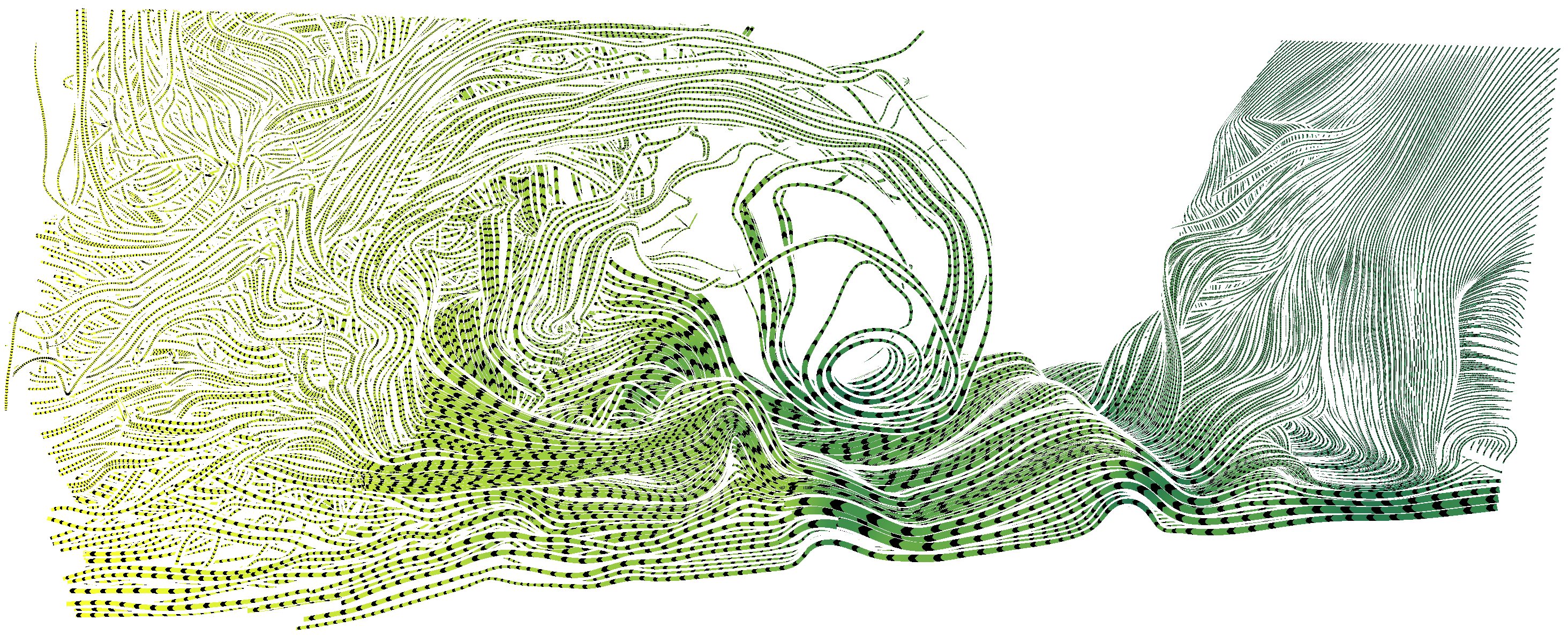}
    \caption{\protect\added{A color map (depicting temperature), combined with a color pattern whose control attribute is the integration time, yielding larger patterns where the velocity is high. The width of the band depends on the local velocity.}}
    \label{fig:colorpattern_velocity_width}
\end{figure}

\added{The color pattern is the final addition to our line style model, of which an application is shown in \autoref{fig:colorpattern_velocity_width}. The base color comes from a yellow-green color map conveying temperature, the alternate color is black. Together, these two colors form a directional pattern, whose length, as in the images shown previously, correlates with the local velocity. This time, however, the width of the centermost band also depends on the velocity, emphasizing the effect.}

\begin{figure}
    \centering
    \includegraphics[width=\columnwidth]{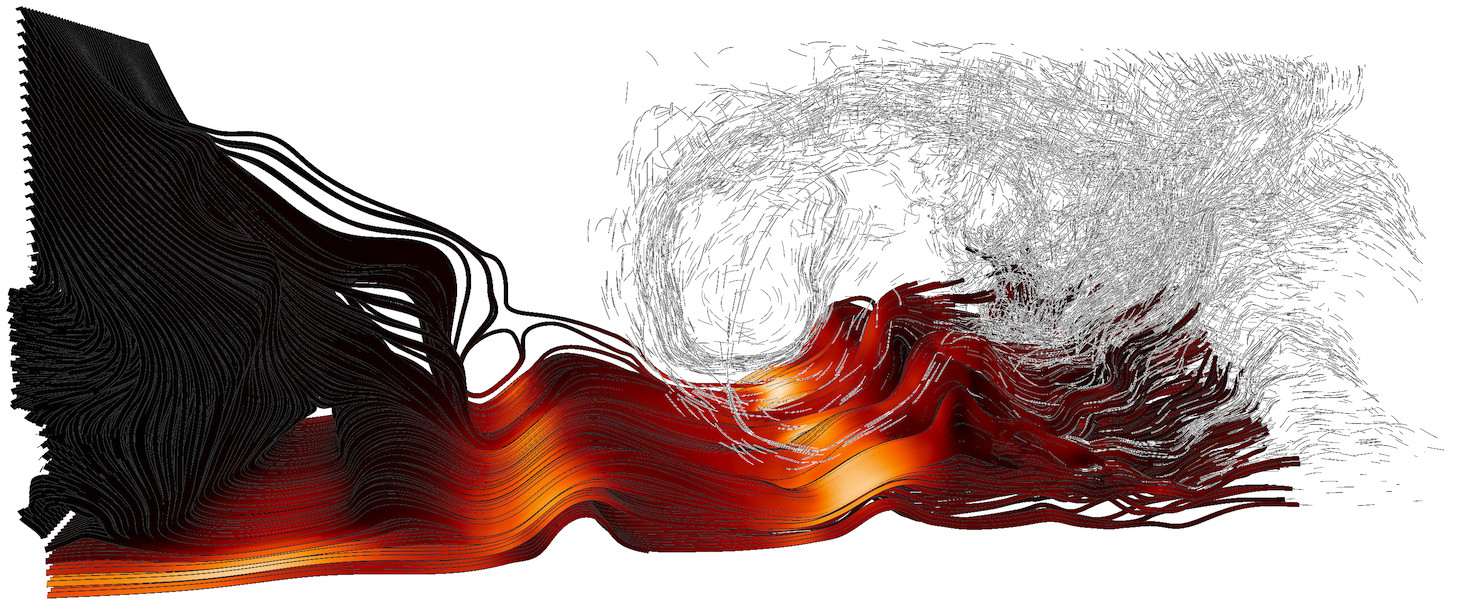}
    \caption{\protect\added{Application of a line style transfer function to emphasize the transition from laminar to turbulent flow while the gray dashed lines provide context.}}
    \label{fig:linestyletransferfunctionexample}
\end{figure}

\added{Finally, in \autoref{fig:linestyletransferfunctionexample} two line styles are shown in one visualization, illustrating the use of a line style transfer function. Here, the visualization or illustration goal is to study the breakup of laminar flow into more turbulent flow. Simply rendering all the lines (as in \autoref{fig:simplebw}) will obscure the transition areas from laminar to turbulent flow; however, by removing the vortex part the context would be lost. For this purpose we defined two line styles for this visualization. One, being the focus of the visualization, is fairly thick and has a color map applied to it (velocity); the other, for providing context, is light gray, thin, and dashed. For this example we use the distance-to-seed-point attribute as the guiding attribute for the line style transfer function, where low values get mapped to the more pronounced linestyle and high values are mapped to the thin dashed lines providing context. Because of the dashing, one can see through the context-providing line, revealing the flow behind it.}

\section{Discussion}
\label{sec:discussion}
As illustrated by the results in the previous section, our parametrization of line styles allows for a wide variety of visual representations of lines, accompanied by visual variables to show additional information about the flow. In this section we discuss further aspects, observations, and limitations of our line style model.

In terms of performance, we find that on a fairly modern graphics card (NVIDIA GeForce GTX 285), we can interactively manipulate the line style parameters whilst displaying fairly large datasets, facilitating the interactive exploration of different visual representations of lines. For reference, the two datasets in \autoref{sec:results} consist of 2500 streamlines (2.5M vertices) and 390 streamlines (250k vertices), respectively. 

\added{The repeating directional patterns in Figures \ref{fig:bwshape}, \ref{fig:whitearrowcolormap}, \ref{fig:blobbyshape01}, and \ref{fig:colorpattern_velocity_width} give a sense of motion to the visualizations. We experimented with this by generating animations where these patterns move along the line, further illustrating the motion (an example is included in the supplementary material). There is, however,  a certain danger in this because it would give the false impression of a steady flow, whereas the datasets we use are a single snapshot of a simulation of unsteady flow. As an experiment to capture the behavior of an \emph{unsteady flow} using our streamline visualization, we generated line visualizations for a number of consecutive snapshots and combined those in a short animation (included in the supplementary material). Although the ends of the streamlines behave somewhat erratically, together, the streamlines seem to capture the behavior of this unsteady flow.}

An additional observation is that in our visualizations where the length of a (shape) pattern depends on the local velocity, the patterns are longer in high velocity areas. Whether this effect is intuitive seems to depend on the people who are asked and the kind of shape being used, as some people correlate high (pattern) frequency with velocity. A related observation is that when the difference in velocity is large, the shape might become too small in low velocity areas, see for example the right side of \autoref{fig:bwshape}.
Other rendering artefacts are possible, for example when (shaped) line strips overlap in a certain way, resulting in oddly shaped patterns. Also, occasionally there are small artefacts when the view-vector is parallel to the line direction, though the effect is minimal and methods exist to remedy this artifact \cite{Stoll:2005:VSL}.
 
%Other perceptual challenges, shape might communicate things not present in the data, though the fact that the bands are mirrored makes it clear what path line the line has. We experimented with adding support for adding `wiggly' lines, but we found that the stylistic local manipulation of the path to support this was not easy to distinguish for the real path, making it less suitable for visualization purposes.
%While experimenting with our line style model, we found that.... interactive seeding.... or other clevering seeding methods.

\section{\ti{Informal Feedback}}
\label{sec:validation}

Finally, we presented our visualization results to a fluid mechanics expert in an informal discussion. In his initial reaction he commented on the ``prettyness'' of the images and continued to state that he finds the visualizations very suitable for illustration purposes (\eg, classroom usage) because they illustrate well-known phenomena very well. Interestingly though, he liked the simple black-and-white visualizations (such as \autoref{fig:simplebw}) best, mainly because of their simplicity and expressive power. 

%\ti{Now about the real validation \dots}

% Remaining possible discussion points
%\begin{itemize}
        %\item Artifacts when line direction is parallel to view-vector 
        %\item Still easy to achieve clutter
        %\item Many parameters
        %\item Transitions in line style transfer functions would be nice.
%\end{itemize}

\section{Conclusion}
We have presented a flexible illustrative line style model for the visualization of streamline datasets. By partitioning line strips into parallel bands whose basic visual properties can be independently controlled, we create a parametrization that allows us to create a broad range of visual styles for line data visualization. This approach is combined with line attribute mapping functions for color and width to facilitate flexible line shapes and means to convey additional information about the flow. Moreover, the line style transfer function we introduced enables emphasis and abstraction by mapping local line attributes to pre-defined line styles. 

Future work includes combining our interactive exploration of line styles with interactive streamline seeding strategies to further improve the interactive exploration of flow datasets for visualization and illustration. Furthermore, we would like to experiment with alternative automatic line style mapping strategies to create more flexible line style transfer functions. 

% Mention rule-based mapping strategies?
% Pathlines \& streaklines in addition to streamlines

\section{Acknowledgements}
We thank Roel Verstappen and F.\ Xavier Trias Miquel for the datasets as well as their discussion and helpful feedback. \ti{Isenberg's work was supported, in part, by a French DIGITEO chair of excellence.}
%-------------------------------------------------------------------------

\bibliographystyle{eg-alpha-doi-doi}

\bibliography{cgf-linestyles}

%-------------------------------------------------------------------------
\newpage

%%%
%%% Figure type 1
%%%

%\begin{figure}[htb]
  %\centering
  %\includegraphics[width=.8\linewidth]{empty}
  %\parbox[t]{.9\columnwidth}{\relax
           %For all figures please keep in mind that you \textbf{must not}
           %use images with transparent background! 
           %}
  %%
  %\caption{\label{fig:firstExample}
           %Here is a sample figure.}
%\end{figure}

%%%
%%% Figure type 3
%%%

%\begin{figure*}[tcb]
  %\centering
  %\mbox{} \hfill
 %\includegraphics[width=.3\linewidth]{empty}
 %\hfill
  %\includegraphics[width=.3\linewidth]{empty}
  %\hfill \mbox{}
  %\caption{\label{fig:ex3}%
           %For publications with color tables (i.e., publications not offering
           %color throughout the paper) please \textbf{observe}: 
           %for the printed version -- and ONLY for the printed
           %version -- color figures have to be placed in the last page.
           %\newline
           %For the electronic version, which will be converted to PDF before
           %making it available electronically, the color images should be
           %embedded within the document. Optionally, other multimedia
           %material may be attached to the electronic version. }
%\end{figure*}

\end{document}